\documentclass[11pt]{article}
\usepackage{amsfonts}
\usepackage{latexsym}
\usepackage{amsmath,amssymb}
\usepackage{verbatim}
\usepackage{setspace}
\usepackage{color}
\usepackage{physics}
\usepackage{tikz}
\usepackage{mathdots}
\usepackage{yhmath}
\usepackage{cancel}
\usepackage{color}
\usepackage{siunitx}
\usepackage{mathrsfs}  
\usepackage{array}
\usepackage{multirow}
\usepackage{amssymb}
\usepackage{gensymb}
\usepackage{tabularx}
\usepackage{hyperref}
\usepackage{booktabs}
\numberwithin{figure}{section}
\usetikzlibrary{fadings}
\usetikzlibrary{patterns}
\usetikzlibrary{shadows.blur}
\usetikzlibrary{shapes}

\usepackage[textheight=9in, textwidth=6.5in, letterpaper]{geometry}

\numberwithin{equation}{section}

\def\ip{${\mathcal I}^+$}

 \def\p{\partial}
 
 \def\bz{{\bar z}}
 
\def\0{{(0)}}
\def\1{{(1)}}
\def\2{{(2)}}

\def\n{\nabla}

\def\<{\langle }
\def\>{\rangle }
\def\[{\left[}
\def\]{\right]}
\def\bw{{\bar w}}

\newcommand{\bea}{\begin{eqnarray}}
\newcommand{\eea}{\end{eqnarray}}
\newcommand{\be}{\begin{equation}}
\newcommand{\ee}{\end{equation}}
\newcommand{\ba}{\begin{align}}
\newcommand{\ea}{\end{align}}

\renewcommand{\O}{\mathcal{O}}

\renewcommand{\epsilon}{\varepsilon}

   \makeatletter
  \let\over=\@@over \let\overwithdelims=\@@overwithdelims
  \let\atop=\@@atop \let\atopwithdelims=\@@atopwithdelims
  \let\above=\@@above \let\abovewithdelims=\@@abovewithdelims
\renewcommand\section{\@startsection {section}{1}{\z@}%
                                   {-3.5ex \@plus -1ex \@minus -.2ex}
                                   {2.3ex \@plus.2ex}%
                                   {\normalfont\large\bfseries}}

\renewcommand\subsection{\@startsection{subsection}{2}{\z@}%
                                     {-3.25ex\@plus -1ex \@minus -.2ex}%
                                     {1.5ex \@plus .2ex}%
                                     {\normalfont\bfseries}}

\begin{document}
\unitlength = 1mm
\ \\
\vskip1cm
\begin{center}

{ \LARGE {\textsc{Deforming Soft Algebras for Gauge Theory}}}

\vspace{0.8cm}
Walker Melton,
Sruthi A. Narayanan, 
and Andrew Strominger

\vspace{1cm}

{\it  Center for the Fundamental Laws of Nature, Harvard University,\\
Cambridge, MA 02138, USA}

\begin{abstract}
Symmetry algebras deriving from towers of soft theorems can be deformed by a short  list of higher-dimension Wilsonian corrections to the effective action.  We study the simplest of  these deformations in gauge theory arising from a massless complex scalar coupled to $F^2$. The soft gauge symmetry `$s$-algebra',  compactly realized as  a higher-spin current algebra acting on the celestial sphere,  is  deformed and enlarged to an associative algebra containing soft scalar generators. This deformed soft algebra is found to be non-abelian even in abelian gauge theory. A two-parameter family of central extensions of  the $s$-subalgebra are generated by shifting and decoupling the scalar generators. It is shown that these central extensions can also be generated by expanding around a certain non-trivial but Lorentz invariant shockwave type background for the scalar field. 
 \end{abstract}
\vspace{0.5cm}

\vspace{1.0cm}

\end{center}

\pagestyle{empty}
\pagestyle{plain}
\newpage
\tableofcontents
\def\gzz{\gamma_{z\bz}}
\def\vx{{\vec x}}
\def\p{\partial}
\def\po{$\cal P_O$}
\def\cN{{\cal N}_\rho^2 }
\def\N{${\cal N}_\rho^2 ~~$}
\def\G{\Gamma}
\def\a{{\alpha}}
\def\b{{\beta}}
\def\g{\gamma}
\def\ch{{\cal H}^+}
\def\hf{{\cal H}}
\def\Im{\mathrm{Im\ }}
\def\bpd{\bar{\partial}}
\def\hbh{{\cal H}_{\rm BH}}
\def\hout{{\cal H}_{\rm OUT}}
\def\ss{\Sigma_S}
\def\D{{\rm \Delta}}
\def \bw {{\bar w}}
\def \bz {{\bar z}}
\def\cS{{\cal S}}
\def\l{\lambda }
\def\d{{\delta}}
\def\n{{\rm SC}}
\def\ip{{\rm cft}}
\def\RR{\mathbb{K}}
\def\i{i^\prime}
\def\A{\mathcal{A}}
\def\zb{\bar{z}}
\def\adz{AdS$_3/\mathbb{Z}$}
\def\sll{$SL(2, \mathbb{R})_L$}
\def\slr{$SL(2, \mathbb{R})_R$}
\renewcommand{\aa}[1]{\left\langle#1\right\rangle}
\pagenumbering{arabic}

\section{Introduction}

Classical minimally-coupled gauge theories and gravity are governed by an infinite number of asymptotic symmetries and conservation laws which constrain both the scattering amplitudes and the time evolution of initial data~\cite{Strominger:2017zoo}. The symmetries are most easily derived from soft theorems and most simply  characterized by generalized higher-spin current algebras\footnote{We consider here the `holomorphic' algebras generated by taking $z\to 0$ with $\bar z$ fixed, where $(z, \bar z)$ are complex coordinates on the celestial sphere.} which act on the  the data on the celestial sphere.  For gravity this algebra has recently been found to contain  the well-known (loop group of the wedge algebra of) $w_{1 +\infty}$, while for gauge theory it is the related `$s$-algebra' \footnote{Originally denoted by  upper case $S$ in \cite{Strominger:2021mtt}.}~\cite{Guevara:2021abz,Strominger:2021mtt}. 

Beyond the classical minimally coupled limit,  one expects this algebra to be deformed in some fashion. A priori there are many possibilities: the symmetries could be anomalous,  the structure constants deformed, central terms generated  or new generators coupled in. However, since these symmetries are associated to the deep infrared they are fully determined by the Wilsonian effective theory. Deformations are expected from both non-minimal Wilsonian corrections to the low energy effective action and quantum loops. While our current understanding is far from complete, work in this  direction includes~\cite{Ball:2021tmb,Ren:2022sws,Costello:2022wso,Costello:2022aol,Bittleston:2022jeq,Ball:2022jac}.

In this paper we consider a Wilsonian deformation  of the $s$-algebra for gauge theory coupled to a complex massless scalar $\phi$.  Only a short list of terms in the Wilsonian effective action can affect the soft theorems or associated symmetries~\cite{Elvang:2017,Pate:2019lpp}. We consider here the deformation induced by the combination of terms $\phi (F^+)^2+\bar \phi (F^-)^2$, where $F^\pm$ are the (anti)-self dual parts of $F$,   appearing in supersymmetric theories as well as the study of Higgs-gluon scattering \cite{Dixon:2004za}. This yields an associative\footnote{Associativity is known for this case but not a priori guaranteed \cite{Ren:2022sws,Costello:2022wso,Costello:2022aol,Bittleston:2022jeq}.} algebra in which  the $s$-algebra is enlarged to include soft scalars as well as soft photons or gluons. We further exhibit a scaling limit in which the scalar generators are shifted by a large amount while the non-minimal coupling is taken to zero, yielding a centrally extended decoupled  $s$-algebra. 

Of special interest is the simple case of photons non-minimally coupled to a massless complex scalar.  Such couplings have appeared in  phenomenological studies of, for example, axions~\cite{Carosi:2013rla}\footnote{As  dark matter candidates axions must have a mass, in which case our considerations apply only above the axion mass scale. Similarly of phenomenological interest  \cite{AristizabalSierra:2021fuc} are neutrino dipole couplings of the form $\bar \nu_R M\sigma_{\lambda\mu}\nu_L F^{\lambda\mu}$ which generate deformations of the soft photon algebra involving soft  neutrino generators. }. In the absence of scalars, all soft photon symmetry generators commute. Interestingly, the non-minimal scalar addition enlarges this abelian algebra with additional soft scalar generators and deforms it to a non-abelian one. 

We  also  consider, in the spirit of~\cite{Fan:2022vbz,Casali:2022fro,Costello:2022wso,Fan:2022kpp,Stieberger:2022zyk,Gonzo:2022tjm},  the expansion of the gauge-scalar theory around a background scalar field vacuum expectation value (vev). If the vev is Lorentz invariant, it will preserve two-dimensional conformal invariance and be dual to a marginal deformation of the boundary CCFT.~\cite{Casali:2022fro,Kapec:2022axw} On the other hand if it breaks translation invariance,  the amplitudes are less singular and the various currents can have non-degenerate two point functions, making the theory easier to study. Such a background vev is provided by $ \phi={ 1 \over x^2}$, which has shockwave type singularities along the light cone of the origin, and is a Lorentz/conformally  invariant cousin of the backgrounds explored in~\cite{Gonzo:2022tjm}. We show that expanding around this vev is equivalent to shifting the soft scalar generators and centrally extends the subleading generators of the $s$-algebra. We further show that, generalizing to a fourth-order kinetic term for the scalar, a similar mechanism imparts a level to the leading (Kac-Moody) generators. This reproduces results of \cite{Costello:2022wso} from an alternate perspective. 

Section~\ref{sec:prelim} contains notation, conventions and brief review. Section~\ref{sec:ymscalar} computes the enlarged and deformed algebra and the central extension of the $s$-algebra. Section~\ref{sec:translationbreak} treats the theory expanded around a Lorentz-invariant scalar vev. 

\section{Preliminaries}
\label{sec:prelim}
In this section we review celestial amplitudes in Klein space and the holomorphic soft gluon symmetry algebra.

\subsection{Celestial Amplitudes  in Klein Space}
On-shell constructions of scattering amplitudes typically employ, implicitly or explicitly,  $(2,2)$ signature Klein space in which left and right spinors as well as the spatial coordinates $(z,\bar z)$ of null infinity  are real and independent. This  allows one  to take holomorphic collinear limits with $z \to 0$ and $\bar z$ fixed, and self-dual gauge fields can be real. More physical $(3,1)$ signature Minkowski space amplitudes may be conveniently obtained by analytic continuation from Klein space.  

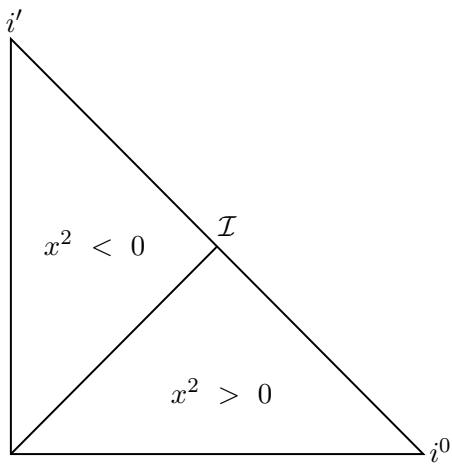
\begin{figure}[htb]
\begin{center}
\tikzset{every picture/.style={line width=0.75pt}} 
\begin{tikzpicture}[x=0.75pt,y=0.75pt,yscale=-0.8,xscale=0.8]
\draw   (100,68.71) -- (360,330.65) -- (100,330.65) -- cycle ;
\draw    (100,330.65) -- (230,199.68) ;
\draw (229,179) node [anchor=north west][inner sep=0.75pt]    {$\mathcal{I} $};
\draw (95,48.35) node [anchor=north west][inner sep=0.75pt]    {$i'$};
\draw (362,318.68) node [anchor=north west][inner sep=0.75pt]    {$i^{0}$};
\draw (119,187.71) node [anchor=north west][inner sep=0.75pt]    {$x^{2} \ < \ 0$};
\draw (199,281.26) node [anchor=north west][inner sep=0.75pt]    {$x^{2} \  >\ 0$};
\end{tikzpicture}
\end{center}
\caption{The toric Penrose diagram for (2,2)-signature Klein space. Each point represents a nondegenerate Lorentzian torus, except the vertical (horizontal) boundary where  the spacelike (timelike) cycle of the torus degenerates. \label{fig:kleinpenrose}}
\end{figure}

The flat Klein space metric is 
\begin{equation}
ds^2 = -dx_0^2+dx_1^2-dx_2^2+dx_3^2
\end{equation}
where $x^\mu$ are the usual Cartesian coordinates. Slices of constant $x^2$ in both the timelike and the spacelike regions of Klein space are geometrically AdS$_3/\mathbb{Z}$. Null infinity takes the form of a Lorentzian torus fibered over a null interval, and the Kleinian equivalent of the celestial sphere is the Lorentzian signature  celestial torus. The toric Penrose diagram of Klein space derived in \cite{Atanasov:2021oyu} is shown in Figure \ref{fig:kleinpenrose}. 
A generic null momentum vector is parameterized  by 
\begin{equation}
q^\mu(z,\bar{z}) =\eta\omega\hat{q}^\mu(z,\bar{z})= \eta\omega(1+z\bar{z},z+\bar{z},z-\bar{z},1-z\bar{z})
\end{equation}
where $z,\bar{z}$ are independent real variables on the celestial torus, $\eta=\pm 1$ denotes an in/out label and $\omega>0$ is the energy.  Using  the basis of Kleinian Pauli matrices
\begin{equation}
\sigma_\mu = \left(\begin{bmatrix} 1 & 0\cr 0 & 1\end{bmatrix}, \begin{bmatrix} 0 & 1\cr 1 & 0\end{bmatrix}, \begin{bmatrix}0 & 1\cr -1 & 0\end{bmatrix}, \begin{bmatrix} -1 & 0\cr 0 & 1\end{bmatrix}\right)
\end{equation}
and letting $q_{\alpha\dot{\alpha}} = (q\cdot \sigma)_{\alpha\dot{\alpha}}=\lambda_\alpha\tilde{\lambda}_{\dot{\alpha}}$ where\footnote{We choose a convention where the spinors are purely imaginary in order to match the spinor products found in the referenced literature.} 
\begin{equation}
\lambda_\alpha = i\eta\sqrt{2\omega}\begin{bmatrix} z\cr 1\end{bmatrix}, \ \ \tilde{\lambda}_{\dot{\alpha}} = -i\sqrt{2\omega}\begin{bmatrix} \bar{z}\cr 1\end{bmatrix}
\end{equation}
we have the following identifications
\begin{equation}
\langle ij\rangle = -2\eta_i\eta_j\sqrt{\omega_i\omega_j}z_{ij}, \ \ [ij] = -2\sqrt{\omega_i\omega_j}\bar{z}_{ij}.
\end{equation}
Throughout this paper we will write amplitudes in terms of these spinor-helicity variables for convenience.

Starting from a massless-particle scattering amplitude in momentum space $A_{a_1\ldots a_n}(q_1,\ldots,q_n)$, where the $a_i$ are gauge group indices, the transform to the celestial conformal primary basis is a Mellin transform with respect to the energies
\begin{equation}
\mathcal{A}\left(1_{\Delta_1,a_1}^{\eta_1},\ldots,n_{\Delta_n,a_n}^{\eta_n}\right) = \left[\prod_{j=1}^n\int_0^\infty d\omega_j\omega_j^{\Delta_j-1}\right]A_{a_1\ldots a_n}\left(\eta_1\omega_1\hat{q}(z_1,\bar{z}_1),\ldots,\eta_n\omega_n\hat{q}(z_n,\bar{z}_n)\right).
\end{equation}  The resulting amplitudes transform as correlation functions of conformal primaries  on the celestial torus. Here $A_{a_1\cdots a_n}$ is implied to contain the momentum conserving delta function. In what follows all celestial amplitudes will be denoted by $\mathcal{A}$ to differentiate them from momentum space amplitudes. When convenient, we will write these amplitudes as correlation functions of the associated primary operators $\O^{a,\eta}_{\Delta}$:
\begin{equation}
\mathcal{A}\left(1_{\Delta_1,a_1}^{\eta_1},\ldots,n_{\Delta_n,a_n}^{\eta_n}\right) = \left\langle \O^{a_1,\eta_1}_{\Delta_1}\cdots \O^{a_n,\eta_n}_{\Delta_n}\right\rangle 
\end{equation}

\subsection{Holomorphic  Soft Algebras }
Celestial amplitudes contain poles where one of the weights approaches a negative integer whose residues are controlled by soft theorems. In the case of gluons, soft poles occur at $\Delta = 1, 0, -1, \ldots$ \cite{Pate:2019mfs}.  For color-ordered gluon amplitudes, the leading soft theorem implies that as $\Delta \to 1$  
\begin{equation}
\lim_{\Delta_1\to 1}(\Delta_1-1)\A\left(1^{\eta_1}_{\Delta_1,+}2^{\eta_2}_{\Delta_2,s_2} \cdots n^{\eta_n}_{\Delta_n,s_n}\right) = -\frac{1}{2}\frac{z_{n2}}{z_{n1}z_{12}}\A\left(2^{\eta_2}_{\Delta_2,s_2}\cdots n^{\eta_n}_{\Delta_n,s_n}\right).
\end{equation}
In the operator language, soft theorems govern insertions of the soft celestial operators
\begin{equation}
R^{k,a}(z,\zb) = \lim_{\varepsilon \to 0}\epsilon\O^{a,+}_{k+\varepsilon},\ k = 1, 0, -1, \ldots 
\end{equation}
where $\O^{a,+}_{\Delta}$ describes a positive helicity gluon. The operator product expansion (OPE) in CCFT can be derived via symmetry constraints or by considering the splitting functions from collinear limits of bulk scattering amplitudes~\cite{Pate:2019lpp,Himwich:2021dau}. Combining the definition of the soft modes with the structure of collinear divergences in Yang-Mills one finds, for gluon scattering without a background, that the positive helicity $\Delta=1$ soft modes form a Kac-Moody algebra with vanishing level~\cite{Strominger:2013lka,He:2015zea}.

The soft operators were shown, in the holomorphic limit, $z\to 0$ with $\bar z$ fixed,  to have the operator product expansions~\cite{Guevara:2021abz}
\begin{equation}
\label{eq:softope}
R^{k,a}(z_1,\zb_1)R^{\ell,b}(z_2,\zb_2) \sim -\frac{if^{ab}_{\ \ c}}{z_{12}}\sum_{m=0}^{1-k}\binom{2-k-\ell-m}{1-\ell}\frac{\zb_{12}^m}{m!}\bar{\partial}^mR^{k+\ell-1,c}(z_2,\zb_2),\ k, \ell = 1, 0, -1, \ldots .
\end{equation}
We can repackage the relevant information contained in the singular part of the soft gluon OPE into an algebra of the modes using the expansion 
\begin{equation}
R^{k,a}(z,\zb) = \sum_{n,m }\frac{R^{k,a}_{n,m}}{z^{n + \frac{k+1}{2}}\zb^{m + \frac{k-1}{2}}},
\end{equation}
where ($m$) $n$ is the (anti)-holomorphic index.
The 2D commutators   can be found through contour integrals of the operator product expansion  \ref{eq:softope}:
\begin{equation}\label{eq:modecontour}
\begin{split}
\left[R^{k,a}_{n,m}, R^{\ell,b}_{n',m'}\right] &= \oint_{|\zb_1| < \epsilon} \frac{d\zb_1}{2\pi i} \zb_1^{m + \frac{k-3}{2}}\oint_{|\bar{z}_2|<\varepsilon} \frac{d\zb_2}{2\pi i}\zb_2^{m' + \frac{\ell-3}{2}} \\
&\oint_{|z_2| < \epsilon}\frac{dz_2}{2\pi i}z_2^{n' + \frac{\ell-1}{2}}\oint_{|z_{12}| < \epsilon}\frac{dz_1}{2\pi i} z_1^{n + \frac{k-1}{2}} R^{k,a}(z_1,\zb_1)R^{\ell,b}(z_2,\zb_2) .
\end{split}
\end{equation}
Defining the wedge modes
\begin{equation}
s^{q,a}_{n,m} = \Gamma(q+m)\Gamma(q-m)R^{3-2q,a}_{n,m},  \ \ \ \frac{k-1}{2}\le  m\le\frac{1-k}{2}  
\end{equation}
we find that the holomorphic soft algebra for gauge theory is  the $s$-algebra~\cite{Strominger:2021mtt}:
\begin{equation}
\left[s^{q,a}_{n,m},s^{p,b}_{n',m'}\right] = -if^{ab}_{\ \ c}s^{p+q-1,c}_{n+n',m+m'}. 
\end{equation}
In what follows we consider deformations of the bulk theory that break translation invariance and generate central extensions as well as deformations arising from dynamical particles in the bulk.

\section{Non-minimal Yang-Mills with a Scalar }\label{sec:ymscalar}
In this section we derive the associative soft algebra  for non-abelian gauge theory non-minimally coupled to a complex scalar. 

\subsection{Deformed algebra}
While the leading soft theorem is exact, coupling gauge theory to massless  scalars can modify the soft theorem at subleading level~\cite{Elvang:2017,Pate:2019lpp}, but only through the short list of  operators that are linear in the scalar and quadratic in the gauge field.  We will see that these modifications deform and enlarge  the soft algebra by including soft scalar generators.  In this paper we study the simplest case  of a complex neutral scalar with action~\cite{Dixon:2004za}
\begin{equation}\label{eq:YMaction}
S = \int d^4x\left(-\partial^\mu\phi\partial_\mu\bar{\phi} -\frac{1}{4}\mbox{Tr}F_{\mu\nu}F^{\mu\nu}  - \frac{\mu}{4}\left(\phi \Tr F^+_{\mu\nu}F^{+\mu\nu} + \bar{\phi}\Tr F^-_{\mu\nu}F^{-\mu\nu}\right)\right).
\end{equation}
In Klein space, $\phi,\bar{\phi}$ become independent real fields. The contributions to the relevant OPEs arising from this interaction are~\cite{Himwich:2021dau}
\begin{equation}
\label{eq:dyndefope}
\begin{split}
R^{k,a}(z_1,\zb_1)R^{\ell,b}(z_2,\zb_2) &\sim -\frac{\mu}{2}\delta^{ab}\frac{\zb_{12}}{z_{12}}\sum_{m=0}^{-k}\binom{-k-\ell-m}{-\ell}\frac{\zb_{12}^m}{m!}\bar{\partial}^m\bar{\phi}^{k+\ell} \\
R^{k,a}(z_1,\zb_1)\phi^k(z_2,\zb_2) &\sim -\frac{\mu}{2}\frac{\zb_{12}}{z_{12}}\sum_{m=0}^{-k}\binom{-1-k-\ell-m}{-\ell-1}\frac{\zb_{12}^m}{m!}\bar{\partial}^m\bar{R}^{k+\ell,a} 
\end{split}
\end{equation}
Here $\bar{\phi}^{k} = \lim_{\epsilon \to 0}\epsilon \bar{\Phi}_{k+\epsilon}$, which is non-vanishing  for $k = 0, -1, \ldots$ and the operator $\bar{\Phi}_{\Delta}$ is dual to a complex scalar. Likewise, $\bar{R}^{k,a} = \lim_{\epsilon\to 0}\epsilon \O_{k+\epsilon}^{a,-}$ is a negative helicity soft gluon. These operators admit the following mode expansions
\begin{equation} 
{\phi}^k= \sum_{n,m} \frac{{\psi}^k_{n,m}}{z^{n+k/2}\zb^{m+k/2}}, \ \ 
\bar{\phi}^k = \sum_{n,m} \frac{\bar{\phi}^k_{n,m}}{z^{n+k/2}\zb^{m+k/2}}, \ \ \bar{R}^{k,a} = \sum_{n,m}\frac{\bar{R}^{k,a}_{n,m}}{z^{n+\frac{k-1}{2}}\zb^{m+\frac{k+1}{2}}}.
\end{equation}
Performing the appropriate contour integrals of the OPEs finally gives us the following deformed algebra 
\begin{eqnarray}\label{eq:gluongluon}
\left[s^{q,a}_{n,m},s^{r,b}_{n',m'}\right] &=& -if^{ab}_cs_{n+n',m+m'}^{r+q-1,c} - \mu\delta^{ab}\left(m'(q-1)-m(r-1)\right)\bar{\sigma}^{q+r-2}_{n+n',m+m'} 
\cr
\left[s^{q,a}_{n,m},\sigma^{r}_{n',m'}\right]&=&-\mu\left(m'(q-1)-m(r-1)\right)\bar{s}^{q+r-2,a}_{n+n',m+m'},
\end{eqnarray}
where we have redefined the modes according to 
\begin{equation}
\sigma^q_{n,m} = \Gamma(q+m)\Gamma(q-m){\phi}^{2-2q}_{n,m}, \ \ \bar{\sigma}^q_{n,m} = \Gamma(q+m)\Gamma(q-m)\bar{\phi}^{2-2q}_{n,m}, \ \ \bar{s}^{q,a}_{n,m} = \Gamma(q+m)\Gamma(q-m)\bar{R}^{1-2q,a}_{n,m}.
\end{equation}
One may check directly from the Jacobi identity that 
this algebra is associative, consistently with the more general analysis of~\cite{Mago:2021wje, Ren:2022sws, Ball:2022jac}. It provides  a simple example of an enlarged and  deformed $s$-algebra. 

\subsection{Centrally Extended $s$-algebra}
Now consider shifting the $\bar \phi$ modes 
\begin{equation} 
\bar{\sigma}^k_{0,0}\rightarrow \bar{\sigma}^k_{0,0}-{\alpha^k \over \mu} 
\end{equation}
and then taking $\mu\rightarrow 0$. The remaining algebra decouples into  associative $\phi$ and $ s$ subalgebras, but the $s$ subalgebra  admits the following deformation
\begin{equation}
\left[s^{q,a}_{n,m},s^{r,b}_{n',m'}\right] = -if^{ab}_cs_{n+n',m+m'}^{r+q-1,c} + \delta^{ab}\left(m'(q-1)-m(r-1)\right)\alpha^{q+r-2}\delta_{n+n'}\delta_{m+m'}. \end{equation}
In the interesting Lorentz/conformally-invariant special case with only $\alpha^1=-c_{3/2}$ nonzero we get 
\begin{eqnarray}\label{eq:centralextended}
[s^{q,a}_{n,m},s^{p,b}_{n',m'}] =  -if^{ab}_{\ \ c}s^{p+q-1,c}_{n+n',m+m'}+ c_{3/2}\delta^{p,3/2}\delta^{q,3/2}\delta^{ab}m\delta_{n+n'}\delta_{m+m'},
\end{eqnarray}
which is a level for $s^{3/2,a}_{n,m}$. 

Shifting $\bar \phi$ is the same as expanding around a bulk background vev whose  explicit form is given below. Hence scalar vevs can generate current algebra levels. A similar shift  in the context of a quartic derivative scalar action generates a level for the leading $s^{1,a}_{n,m}$  as in \cite{Costello:2022wso}. We now  construct bulk theories whose amplitudes realize the algebra~\eqref{eq:centralextended}.

\section{Translation Non-Invariant Backgrounds }\label{sec:translationbreak}
In this section we consider the  scalar-Yang-Mills theory~\eqref{eq:YMaction} in  a non-translationally invariant scalar background. 

Consider, first,  backgrounds\footnote{We consider only a background for the holomorphic $\phi$; the case where $\bar{\phi}$ can be found by parity-conjugating the results of this work. } $\phi_B(x)$  as in \cite{Casali:2022fro} (without a mass) 
satisfying $\square\phi_B = 0$ (or $\square^2\phi_B = 0$ for a scalar with a fourth-order kinetic term). The action ~\eqref{eq:YMaction} expanded around such a background acquires an extra term 
\begin{equation}
\delta S_B= -\int d^4x\frac{\mu}{4}\phi_B(x)\Tr F^+_{\mu\nu}F^{+\mu\nu}.
\end{equation}
In the case that $\phi_B$ is Lorentz invariant, computing amplitudes around such a linearized  on-shell background is equivalent to deforming the boundary celestial conformal field theory by a dimension two marginal operator~\cite{Casali:2022fro,Kapec:2022axw}.

In the following we are interested in backgrounds that generate central extensions to the gauge symmetry algebra. 
As will be seen shortly, this turns out to require that the massless wave equation has a source. The source $J$ could, for example, be nonzero $(F^-)^{2}$, D-branes as in~\cite{Costello:2022bur} or a quantum  violation of the classical equation of motion from contact terms with an operator insertion. We remain agnostic about the precise origin and simply add an explicit source term to the action
\begin{equation}
\delta S_S = -\int d^4x\left[J(x)\bar{\phi}(x) \right].
\end{equation}
By examining Feynman diagrams with this action, one finds that this is equivalent to computing amplitudes around a background that solves $\square\phi_B = J$
\begin{equation}
\phi_B(x) = \int d^4x'J(x)\Delta(x,x')
\end{equation}
where $\Delta$ is the Feynman propagator of the massless scalar. In momentum space, the effective background can be written as\footnote{One could add a homogeneous contribution to this background, but such a term will not deform the soft algebra.}
\begin{equation}
\phi_B(p) = \Delta(-p)J(p).
\end{equation}
This leads to, at linear order in $\phi_B$, the $n$-gluon amplitudes
\begin{equation}
A_n^\phi(1^+\cdots n^+) = \int d^4p_0\phi_B(p_0)A_{n+1}(0^01^+\cdots n^+)
\end{equation}
If $\phi_B(p)$ depends only on $p^2$, the resulting amplitudes will be Lorentz but not translation invariant and hence correspond to a marginal deformation of the boundary CCFT.  The amplitude $A^\phi_{n+1}(0^01^+\cdots n^+)$ has been studied in the context of Higgs-gluon scattering~\cite{Dixon:2004za}, and the color ordered amplitude takes the simple form 
\begin{equation}
A_{n+1}(0^01^+\cdots n^+) = \frac{\mu (p_0^2)^2}{\langle 12 \rangle \langle 23 \rangle \cdots \langle n1\rangle}\delta^{(4)}(p_0 + p_1 + \cdots + p_n).
\end{equation}
The celestial amplitude of $n$ positive helicity gluons is then the  Mellin transform of $A^\phi_n(1^+\cdots n^+)$. We now  compute these amplitudes in the specific case where all external particles are taken to be soft positive helicity gluons.

\subsection{All-Soft Amplitudes in a  Background }
\label{sec:allsoftamps}
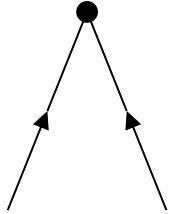
\begin{figure}[h]
\begin{center}

\tikzset{every picture/.style={line width=0.75pt}} 

\begin{tikzpicture}[x=0.75pt,y=0.75pt,yscale=-1,xscale=1]

\draw    (98.89,112.79) -- (80,160) ;
\draw [shift={(100,110)}, rotate = 111.8] [fill={rgb, 255:red, 0; green, 0; blue, 0 }  ][line width=0.08]  [draw opacity=0] (8.93,-4.29) -- (0,0) -- (8.93,4.29) -- cycle    ;
\draw    (160,160) -- (141.11,112.79) ;
\draw [shift={(140,110)}, rotate = 68.2] [fill={rgb, 255:red, 0; green, 0; blue, 0 }  ][line width=0.08]  [draw opacity=0] (8.93,-4.29) -- (0,0) -- (8.93,4.29) -- cycle    ;
\draw    (120,60) -- (100,110) ;
\draw    (120,60) -- (140,110) ;
\draw  [fill={rgb, 255:red, 0; green, 0; blue, 0 }  ,fill opacity=1 ] (115,60) .. controls (115,57.24) and (117.24,55) .. (120,55) .. controls (122.76,55) and (125,57.24) .. (125,60) .. controls (125,62.76) and (122.76,65) .. (120,65) .. controls (117.24,65) and (115,62.76) .. (115,60) -- cycle ;

\end{tikzpicture}
\end{center}
\caption{A two-gluon scattering process at first order in the background. \label{fig:twogluon}}
\end{figure}

For all particles ingoing, these celestial amplitudes describe $n$ gluons scattering into nothing by interacting with a background, such as the two-gluon scattering amplitude illustrated in Figure \ref{fig:twogluon}
As such, the background deformation will deform the soft gluon algebra by giving a non-zero two point function to soft gluons, describing a process where two positive-helicity soft gluons annihilate in the presence of the background. We consider the simplest case of the two-point function first. One finds \footnote{For notational simplicity we consider only ingoing particles and drop the ingoing/outgoing label $\eta_j$.}
\begin{equation}
\label{eq:twopoint}
\left\langle R^{k,a}(z_1,\zb_1)R^{\ell,b}(z_2,\zb_2) \right\rangle_\phi = \Tr(T^aT^b) \lim_{\epsilon_1,\epsilon_2\to 0}\epsilon_1\epsilon_2\mathcal{A}^\phi\left(1^+_{k+\epsilon_1}2^+_{\ell+\epsilon_2}\right)  = \frac{2\mu\delta_{k\ell}\delta^{ab}a_{-1-k}}{(-4)^kz_{12}^2|z_{12}|^{2(k-1)}}
\end{equation}
where we have assumed that $\phi_B$ is damped in the UV and that, near $p = 0$, $\phi_B(p)$ has the expansion 
\begin{equation}
\phi_B(p) = \sum_k a_k(p^2)^k.
\end{equation}
The details of this computation can be found in Appendix~\ref{app:celestialamplitudes}. 

It should be noted that, in the presence of an effective background with nonzero $a_{-1-k}$, equation \ref{eq:twopoint} implies that the operator $R^{k,a}$ creates states of finite norm when acting on the vacuum. This result can be generalized to higher point functions. As an example, the three point function is computed explicitly in Appendix~\ref{app:celestialamplitudes}.

The deformations of soft gluon amplitudes will always contain a factor of $\mu a_k$. Of special interest to us is the limit  $\mu \to 0 $ with $\mu a_k$ fixed, in which scalars decouple from the soft gluons while the latter acquire central extensions. 
\subsubsection{Leading Soft Gluons}
For conformal dimension $k=1$ we get the two and three point functions between leading soft currents which are 
\begin{equation}
\begin{split}
\left\langle R^{1,a}(z_1)R^{1,b}(z_2)\right\rangle_\phi &= \frac{c_1\delta^{ab}}{z_{12}^2}, \ \ \\
\left\langle R^{1,a}(z_1)R^{1,b}(z_2)R^{1,c}(z_3)\right\rangle_\phi &= -\frac{c_1 f^{abc}}{z_{12}z_{23}z_{31}}.
\end{split}
\end{equation}
where 
\begin{equation}
c_1 = -\frac{\mu a_{-2}}{2} = -\frac{\mu}{2}\int d^4x\square^2\phi_B(x)  .
\end{equation}
As an example, consider the choice $\phi_B(x) = \alpha\log |x|^2$, which  solves the equations of motion for a scalar with kinetic term $\bar \phi\square^2\phi$ with a $\delta$-function source at the origin. 
This creates a non-vanishing soft two-point function with $c_1 = 8\pi^2\mu \alpha$, in agreement with the topological field theory analysis of  \cite{Costello:2022wso}.
 
 A simple generalization of the above calculation yields the $n$-point function of positive-helicity leading soft gluons:
\begin{equation}
\left\langle R^{1,a_1}(z_1)\cdots R^{1,a_n}(z_n)\right\rangle_\phi = -\frac{c_1(-1)^n \mbox{Tr}[T^{a_1}\cdots T^{a_n}]}{2^{n-1}z_{12}\cdots z_{n1}} + \cdots
\end{equation}
where $\cdots$ include other color orderings and terms that are higher-trace. This takes the form of a correlation function of $n$ holomorphic currents in a WZW model~\cite{DiFrancesco:639405} of level $k = -\frac{\mu a_{-2}}{2}$. This is reminiscent of results in~\cite{Costello:2022wso,Fan:2022kpp,Costello:2022bur,Stieberger:2022zyk}.

\subsubsection{Subleading Soft Gluons}
Similarly, the conformal dimension $k=0$ soft gluons, which govern the subleading soft theorem, have two point function
\begin{equation}
\langle R^{0,a}(z_1,\zb_1) R^{0,b}(z_2,\zb_2) \rangle_\phi = c_{3/2}\delta^{ab}\frac{\zb_{12}}{z_{12}}
\end{equation}
where 
\begin{equation}
c_{3/2} = 2\mu a_{-1} = -2\mu\int d^4x\square\phi_B(x) = -2\mu\int d^4xJ(x)
\end{equation}
The effective background $\phi_B(x) = \frac{\alpha}{x^2}$, with $i\epsilon$ prescription  solving  the massless Klein-Gordon equation with  $\delta$-function source at the origin, gives the subleading soft gluon a two point function with $c_{3/2} = 8\pi^2\mu\alpha$. Note that adding a solution to the homogeneous equation of motion $\square^2\phi = 0$ does not contribute to $c_1$.

Lorentz invariance forces the two-point function $\langle R^{k,a}R^{k,b}\rangle$ to be non-singular as $z_{12} \to 0$ for $k \le 0$, so these are the only two deformations that will modify the soft algebra.

\subsection{Background Induced Central Extensions}
Observation of the scattering amplitudes given in subsection~\ref{sec:allsoftamps} reveals a simple structure: at linear order in $\mu a_k$, the background deforms the soft OPE by adding a term proportional to the identity 
\begin{eqnarray}
R^{k,a}(z_1,\zb_1)R^{\ell,b}(z_2,\zb_2) &\sim & -\frac{if^{ab}_{\ \ c}}{z_{12}}\sum_{m=0}^{1-k}\binom{2-k-\ell-m}{1-\ell}\frac{\zb_{12}^m}{m!}\bar{\partial}^mR^{k+\ell-1,c}(z_2,\zb_2) \cr
& - &\frac{\mu}{2}\delta^{ab}\frac{\zb_{12}}{z_{12}}\sum_{m=0}^{-k}\binom{-k-\ell-m}{-\ell}\frac{\zb_{12}^m}{m!}\bar{\partial}^m\bar{\phi}^{k+\ell} + \frac{2\mu\delta_{k\ell}\delta^{ab}a_{-1-k}}{(-4)^kz_{12}^2|z_{12}|^{2(k-1)}} ~~~
\end{eqnarray}
This deforms the soft algebra when the identity term is singular as $z_{12} \to 0$, which occurs at leading and subleading level. We can explicitly compute this deformation by performing the appropriate contour integrals outlined in Equation~\eqref{eq:modecontour}. The centrally extended algebra in the $\mu\to 0$ 
 and $\mu a_k$ fixed limit is  
\begin{equation}
[s^{q,a}_{n,m},s^{p,b}_{n',m'}]=  -if^{ab}_{\ \ c}s^{p+q-1,c}_{n+n',m+m'}+
c_1\delta^{p1}\delta^{q1}\delta^{ab}n\delta_{n+n'}\delta_{m+m'}  + c_{3/2}\delta^{p,3/2}\delta^{q,3/2}\delta^{ab}m\delta_{n+n'}\delta_{m+m'}.
\end{equation}

\section{Discussion}

Deformations to soft algebras in celestial CFTs have recently been an area of much interest. In this work, we have shown that interesting deformations of soft algebras can arise from deforming celestial CFTs by computing scattering amplitudes with scalar sources and a non-trivial background. To first order in the source, these deformations centrally extend the soft algebra, or, in a complementary picture, give nonzero vacuum expectation values to weight 0 scalar boundary operators as expected from the analogy with conformal perturbation theory, provided that the source arises from dynamical element of the theory. Interestingly, on-shell backgrounds (with vanishing $J(x)$) do not deform the soft algebra.

Along the way, we have replicated an interesting result from \cite{Costello:2022wso}, where correlation functions of chiral WZW models appeared to describe correlation functions of positive-helicity leading soft gluons computed around a logarithmic background. The derivation presented in this paper is a direct computation by taking the soft limit of a celestial amplitude, rather than an indirect gauge-invariance argument, providing an interesting check that these two perspectives on 4D/2D holography are equivalent, and extends their derivation to correlation functions of subleading soft gluons and more general sources.

This work also invites us to investigate soft scattering with sources as a testbed for celestial duality. Because the two-point functions of soft gluons in a background can be nonvanishing, soft gluons create states of finite norm when computed around a non-trivial background, and at leading order, have correlation functions reminiscent of chiral WZW models. It would be interesting to see whether this correspondence can be extended to higher orders in the background to include negative helicity gluons and dynamic scalar while maintaining a duality with a known boundary theory.\footnote{Along these lines   an exact duality between scattering of plane-wave-like states in a WZW model on the asymptotically flat Burns metric and a 2D current algebra has been proposed, where the Burns metric itself plays the role of the logarithmic background and allows non-vanishing soft two point functions \cite{Costello:2022bur}.}

We expect to follow up with the somewhat analogous results in gravity in the near future.

\section*{Acknowledgements}
We are grateful to Adam Ball, Luca Ciambelli, Kevin Costello, Atul Sharma, Natalie Paquette, Sabrina Pasterski,  Akshay Srikant and Tom Taylor for useful conversations. This work was supported in part by NSF grant PHY-2207659 and NSF GRFP grant DGE1745303. 

\appendix
\section{Computation of Correlators in a Background}\label{app:celestialamplitudes}
In this appendix we provide details for the computation of two and three point functions of soft operators in Yang-Mills with a scalar background. The celestial $n$-point function in a background can be computed as follows
\begin{equation}
\begin{split}
\A_n^\phi(1^{+,\eta_1}_{\Delta_1}\cdots n^{+,\eta_n}_{\Delta_n}) &=\frac{\mu(-1)^n}{2^nz_{12}\cdots z_{n1}} \int_0^\infty d\omega_j\omega_j^{\Delta_j-2} \int d^4p_0 (p_0^2)^2\phi_B(p_0)\delta^{(4)}(p_0 + \eta_1\omega_1q(z_1,\zb_1) + \cdots) \\
&= \frac{\mu(-1)^n}{2^nz_{12}\cdots z_{n1}}\int_0^\infty d\omega_j\omega_j^{\Delta_j-1}(p^2)^2\phi_B(p)\bigg|_{p = -\eta_1\omega_1q(z_1,\zb_1) - \cdots - \eta_n\omega_nq(z_n,\zb_n)}.
\end{split}
\end{equation} Specifying to $n=2$, the two point function of soft operators is given by the Mellin transform combined with the appropriate soft limits of the above
\begin{eqnarray}
\A^\phi\left(1^+_k2^+_\ell\right) & = & \lim_{\epsilon_1,\epsilon_2\to 0}\epsilon_1\epsilon_2 \A\left(1^+_{k+\epsilon_1}2^+_{\ell+\epsilon_2}\right)\cr
& = &  -\frac{\mu}{4z_{12}^2}\int_0^\infty d\omega_1\left(\epsilon_1\omega_1^{k-2+\epsilon_1}\right)d\omega_2\left(\epsilon_2\omega_2^{\ell-2+\epsilon_2}\right)\left(\left(\omega_1q(z_1,\zb_1)+\omega_2q_2(z_2,\zb_2)\right)^2\right)^2 \cr
 &\times &\phi_B\left(-\omega_1q(z_1,\zb_1)-\omega_2q(z_2,\zb_2)\right).
\end{eqnarray}
Assuming that $\phi$ is damped in the UV, we can use the following property of the Dirac delta function
\begin{equation}
   \frac{2(-1)^n}{n!} \delta^{(n)}(x) = \lim_{\epsilon \to 0} \epsilon |x|^{-1-n+\epsilon}.
\end{equation}
Additionally assuming that, near $p = 0$, $\phi_B(p)$ has the expansion
\begin{equation}
\phi_B(p) = \sum_k a_k(p^2)^k,
\end{equation}
we can substitute back into our expression to obtain
\begin{equation}
\begin{split}
\mathcal{A}^\phi\left(1^+_k2^+_\ell\right) &= -\frac{\mu}{4z_{12}^2}\int d\omega_1d\omega_2\frac{(-1)^{2-k-\ell}\delta^{(1-k)}(\omega_1)\delta^{(1-\ell)}(\omega_2)}{(1-k)!(1-\ell)!}\sum_ja_j\left(-4\omega_1\omega_2|z_{12}|^2\right)^{2+j} \\
 &= \frac{\mu\delta_{k\ell}a_{-1-k}}{(-4)^kz_{12}^2}
\end{split}
\end{equation}
In fact, the three point function can be computed in a similar fashion. For generic soft gluons we find, defining $\beta = k_1 + k_2 + k_3-3$,
\begin{equation}
\begin{split}
\A^\phi(1^+_{k_1}2^+_{k_2}3^+_{k_3}) &= \lim_{\epsilon_j\to 0}\epsilon_1\epsilon_2\epsilon_3\A(1^+_{k_1+\epsilon_1}2^+_{k_2+\epsilon_2}3^+_{k_3+\epsilon_3}) \\
&=-\lim_{\epsilon_j\to 0} \frac{\mu}{8z_{12}z_{23}z_{31}}\int \prod_jd\omega_j\epsilon_j\omega_j^{k_j-2+\epsilon_j}(p^2)^2\phi_B(p)\bigg|_{p = -\omega_1q_1 - \omega_2q_2 - \omega_3q_3} \\
&=- \frac{\mu}{8z_{12}z_{23}z_{31}}\int d\omega_1d\omega_2d\omega_3\frac{(-1)^{-\beta}\delta^{(1-k_1)}(\omega_1)\delta^{(1-k_2)}(\omega_2)\delta^{(1-k_3)}(\omega_3)}{(1-k_1)!(1-k_2)!(1-k_3)!}\\
& \times  (p^2)^2\phi_B(p)\bigg|_{p = -\omega_1q_1 - \omega_2q_2 - \omega_3q_3}
\end{split}
\end{equation}
The effect of the soft limit is to pick out the coefficient of $\omega_1^{1-k_1}\omega_2^{1-k_2}\omega_3^{1-k_3}$ in the Laurent expansion of $p^4\phi_B(p)|_{-\omega_1q_1-\omega_2q_2-\omega_3q_3}$. Therefore, our three-point function takes the form 
\begin{equation}
\begin{split}
\A^\phi(1^+_{k_1}2^+_{k_2}3^+_{k_3}) &= \frac{(-1)^{1-\beta}(-4)^{-\beta/2}a_{-2-\beta/2}}{8z_{12}z_{23}z_{31}}\int d\omega_1d\omega_2d\omega_3\frac{\delta^{(1-k_1)}(\omega_1)\delta^{(1-k_2)}(\omega_2)\delta^{(1-k_3)}(\omega_3)}{(1-k_1)!(1-k_2)!(1-k_3)!}\\
 &\times \binom{-\frac{\beta}{2}}{\frac{k_1-k_2-k_3+1}{2},\frac{-k_1+k_2-k_3+1}{2},\frac{-k_1-k_2+k_3+1}{2}}\omega_1^{1-k_1}\omega_2^{1-k_2}\omega_3^{1-k_3}\\
 &\times \frac{1}{|z_{12}|^{k_1+k_2-k_3-1}|z_{23}|^{k_2+k_3-k_1-1}|z_{13}|^{k_1+k_3-k_2-1}} \\
&= -\frac{a_{-2-\beta/2}}{(-1)^{3\beta/2}2^{\beta+3}z_{12}z_{23}z_{31}}\binom{-\frac{\beta}{2}}{\frac{k_1-k_2-k_3+1}{2},\frac{-k_1+k_2-k_3+1}{2},\frac{-k_1-k_2+k_3+1}{2}} \\
&\times \frac{1}{|z_{12}|^{k_1+k_2-k_3-1}|z_{23}|^{k_2+k_3-k_1-1}|z_{13}|^{k_1+k_3-k_2-1}}
\end{split}
\end{equation}
This is nonvanishing if $k_i + k_j - k_p \le 1$ for $i \ne j \ne p$ and $\beta$ even.

\bibliographystyle{utphys}
\bibliography{cpb}

\providecommand{\href}[2]{#2}\begingroup\raggedright\begin{thebibliography}{10}

\bibitem{Strominger:2017zoo}
A.~Strominger, {\em Lectures on the Infrared Structure of Gravity and Gauge
  Theory}.
\newblock Princeton University Press, 2018.
\newblock \href{http://arxiv.org/abs/1703.05448}{{\ttfamily arXiv:1703.05448
  [hep-th]}}.

\bibitem{Strominger:2021mtt}
A.~Strominger, ``{$w_{1+\infty}$ Algebra and the Celestial Sphere: Infinite
  Towers of Soft Graviton, Photon, and Gluon Symmetries},''
  \href{http://dx.doi.org/10.1103/PhysRevLett.127.221601}{{\em Phys. Rev.
  Lett.} {\bfseries 127} no.~22, (2021) 221601}.

\bibitem{Guevara:2021abz}
A.~Guevara, E.~Himwich, M.~Pate, and A.~Strominger, ``{Holographic Symmetry
  Algebras for Gauge Theory and Gravity},''
  \href{http://arxiv.org/abs/2103.03961}{{\ttfamily arXiv:2103.03961
  [hep-th]}}.

\bibitem{Ball:2021tmb}
A.~Ball, S.~A. Narayanan, J.~Salzer, and A.~Strominger, ``{Perturbatively exact
  w$_{1+\infty}$ asymptotic symmetry of quantum self-dual gravity},''
  \href{http://dx.doi.org/10.1007/JHEP01(2022)114}{{\em JHEP} {\bfseries 01}
  (2022) 114}, \href{http://arxiv.org/abs/2111.10392}{{\ttfamily
  arXiv:2111.10392 [hep-th]}}.

\bibitem{Ren:2022sws}
L.~Ren, M.~Spradlin, A.~Yelleshpur~Srikant, and A.~Volovich, ``{On effective
  field theories with celestial duals},''
  \href{http://dx.doi.org/10.1007/JHEP08(2022)251}{{\em JHEP} {\bfseries 08}
  (2022) 251}, \href{http://arxiv.org/abs/2206.08322}{{\ttfamily
  arXiv:2206.08322 [hep-th]}}.

\bibitem{Costello:2022wso}
K.~Costello and N.~M. Paquette, ``{Celestial holography meets twisted
  holography: 4d amplitudes from chiral correlators},''
  \href{http://arxiv.org/abs/2201.02595}{{\ttfamily arXiv:2201.02595
  [hep-th]}}.

\bibitem{Costello:2022aol}
K.~Costello and N.~M. Paquette, ``{On the associativity of one-loop corrections
  to the celestial OPE},'' \href{http://arxiv.org/abs/2204.05301}{{\ttfamily
  arXiv:2204.05301 [hep-th]}}.

\bibitem{Bittleston:2022jeq}
R.~Bittleston, ``{On the associativity of 1-loop corrections to the celestial
  operator product in gravity},''
  \href{http://arxiv.org/abs/2211.06417}{{\ttfamily arXiv:2211.06417
  [hep-th]}}.

\bibitem{Ball:2022jac}
A.~Ball, ``{Celestial Locality and the Jacobi Identity},''
  \href{http://arxiv.org/abs/2211.09151}{{\ttfamily arXiv:2211.09151
  [hep-th]}}.

\bibitem{Elvang:2017}
H.~Elvang, C.~R.~T. Jones, and S.~G. Naculich, ``Soft photon and graviton
  theorems in effective field theory,''
  \href{http://dx.doi.org/10.1103/PhysRevLett.118.231601}{{\em Phys. Rev.
  Lett.} {\bfseries 118} (Jun, 2017) 231601}.
  \url{https://link.aps.org/doi/10.1103/PhysRevLett.118.231601}.

\bibitem{Pate:2019lpp}
M.~Pate, A.-M. Raclariu, A.~Strominger, and E.~Y. Yuan, ``{Celestial Operator
  Products of Gluons and Gravitons},''
  \href{http://arxiv.org/abs/1910.07424}{{\ttfamily arXiv:1910.07424
  [hep-th]}}.

\bibitem{Dixon:2004za}
L.~J. Dixon, E.~W.~N. Glover, and V.~V. Khoze, ``{MHV rules for Higgs plus
  multi-gluon amplitudes},''
  \href{http://dx.doi.org/10.1088/1126-6708/2004/12/015}{{\em JHEP} {\bfseries
  12} (2004) 015}, \href{http://arxiv.org/abs/hep-th/0411092}{{\ttfamily
  arXiv:hep-th/0411092}}.

\bibitem{Carosi:2013rla}
G.~Carosi, A.~Friedland, M.~Giannotti, M.~J. Pivovaroff, J.~Ruz, and J.~K.
  Vogel, ``{Probing the axion-photon coupling: phenomenological and
  experimental perspectives. A snowmass white paper},'' in {\em {Community
  Summer Study 2013}: {Snowmass on the Mississippi}}.
\newblock 9, 2013.
\newblock \href{http://arxiv.org/abs/1309.7035}{{\ttfamily arXiv:1309.7035
  [hep-ph]}}.

\bibitem{AristizabalSierra:2021fuc}
D.~Aristizabal~Sierra, O.~G. Miranda, D.~K. Papoulias, and G.~S. Garcia,
  ``{Neutrino magnetic and electric dipole moments: From measurements to
  parameter space},'' \href{http://dx.doi.org/10.1103/PhysRevD.105.035027}{{\em
  Phys. Rev. D} {\bfseries 105} no.~3, (2022) 035027},
  \href{http://arxiv.org/abs/2112.12817}{{\ttfamily arXiv:2112.12817
  [hep-ph]}}.

\bibitem{Fan:2022vbz}
W.~Fan, A.~Fotopoulos, S.~Stieberger, T.~R. Taylor, and B.~Zhu, ``{Elements of
  celestial conformal field theory},''
  \href{http://dx.doi.org/10.1007/JHEP08(2022)213}{{\em JHEP} {\bfseries 08}
  (2022) 213}, \href{http://arxiv.org/abs/2202.08288}{{\ttfamily
  arXiv:2202.08288 [hep-th]}}.

\bibitem{Casali:2022fro}
E.~Casali, W.~Melton, and A.~Strominger, ``{Celestial Amplitudes as AdS-Witten
  Diagrams},'' \href{http://arxiv.org/abs/2204.10249}{{\ttfamily
  arXiv:2204.10249 [hep-th]}}.

\bibitem{Fan:2022kpp}
W.~Fan, A.~Fotopoulos, S.~Stieberger, T.~R. Taylor, and B.~Zhu, ``{Celestial
  Yang-Mills Amplitudes and D=4 Conformal Blocks},''
  \href{http://arxiv.org/abs/2206.08979}{{\ttfamily arXiv:2206.08979
  [hep-th]}}.

\bibitem{Stieberger:2022zyk}
S.~Stieberger, T.~R. Taylor, and B.~Zhu, ``{Celestial Liouville Theory for
  Yang-Mills Amplitudes},'' \href{http://arxiv.org/abs/2209.02724}{{\ttfamily
  arXiv:2209.02724 [hep-th]}}.

\bibitem{Gonzo:2022tjm}
R.~Gonzo, T.~McLoughlin, and A.~Puhm, ``{Celestial holography on Kerr-Schild
  backgrounds},'' \href{http://dx.doi.org/10.1007/JHEP10(2022)073}{{\em JHEP}
  {\bfseries 10} (2022) 073}, \href{http://arxiv.org/abs/2207.13719}{{\ttfamily
  arXiv:2207.13719 [hep-th]}}.

\bibitem{Kapec:2022axw}
D.~Kapec, Y.~T.~A. Law, and S.~A. Narayanan, ``{Soft Scalars and the Geometry
  of the Space of Celestial CFTs},''
  \href{http://arxiv.org/abs/2205.10935}{{\ttfamily arXiv:2205.10935
  [hep-th]}}.

\bibitem{Atanasov:2021oyu}
A.~Atanasov, A.~Ball, W.~Melton, A.-M. Raclariu, and A.~Strominger, ``{$(2,2)$
  Scattering and the Celestial Torus},''
  \href{http://arxiv.org/abs/2101.09591}{{\ttfamily arXiv:2101.09591
  [hep-th]}}.

\bibitem{Pate:2019mfs}
M.~Pate, A.-M. Raclariu, and A.~Strominger, ``{Conformally Soft Theorem in
  Gauge Theory},'' \href{http://dx.doi.org/10.1103/PhysRevD.100.085017}{{\em
  Phys. Rev. D} {\bfseries 100} no.~8, (2019) 085017},
  \href{http://arxiv.org/abs/1904.10831}{{\ttfamily arXiv:1904.10831
  [hep-th]}}.

\bibitem{Himwich:2021dau}
E.~Himwich, M.~Pate, and K.~Singh, ``{Celestial operator product expansions and
  w(1+infinity) symmetry for all spins},''
  \href{http://dx.doi.org/10.1007/JHEP01(2022)080}{{\em JHEP} {\bfseries 01}
  (2022) 080}, \href{http://arxiv.org/abs/2108.07763}{{\ttfamily
  arXiv:2108.07763 [hep-th]}}.

\bibitem{Strominger:2013lka}
A.~Strominger, ``{Asymptotic Symmetries of Yang-Mills Theory},''
  \href{http://dx.doi.org/10.1007/JHEP07(2014)151}{{\em JHEP} {\bfseries 07}
  (2014) 151}, \href{http://arxiv.org/abs/1308.0589}{{\ttfamily arXiv:1308.0589
  [hep-th]}}.

\bibitem{He:2015zea}
T.~He, P.~Mitra, and A.~Strominger, ``{2D Kac-Moody Symmetry of 4D Yang-Mills
  Theory},'' \href{http://dx.doi.org/10.1007/JHEP10(2016)137}{{\em JHEP}
  {\bfseries 10} (2016) 137}, \href{http://arxiv.org/abs/1503.02663}{{\ttfamily
  arXiv:1503.02663 [hep-th]}}.

\bibitem{Mago:2021wje}
J.~Mago, L.~Ren, A.~Y. Srikant, and A.~Volovich, ``{Deformed $w_{1+\infty}$
  Algebras in the Celestial CFT},''
  \href{http://arxiv.org/abs/2111.11356}{{\ttfamily arXiv:2111.11356
  [hep-th]}}.

\bibitem{Costello:2022bur}
K.~Costello, N.~M. Paquette, and A.~Sharma, ``{Top-down holography in an
  asymptotically flat spacetime},''
  \href{http://arxiv.org/abs/2208.14233}{{\ttfamily arXiv:2208.14233
  [hep-th]}}.

\bibitem{DiFrancesco:639405}
P.~Di~Francesco, P.~Mathieu, and D.~Sénéchal,
  \href{http://dx.doi.org/10.1007/978-1-4612-2256-9}{{\em {Conformal field
  theory}}}.
\newblock Graduate texts in contemporary physics. Springer, New York, NY, 1997.
\newblock \url{https://cds.cern.ch/record/639405}.

\end{thebibliography}\endgroup

\end{document}